Virgin curve outside the envelope in isothermal M-H: a Comment on Nayak et al PRL 110 (2013) 127204

P Chaddah
UGC-DAE Consortium for Scientific Research,
Indore 452001
India.

The ease of experimentally varying magnetic field H (transmitted in vacuum), in comparison to that of varying pressure (transmitted through a material medium) in conjunction with varying temperature, has made popular the study of magnetic first order transitions to understand metastabilities associated with first order transitions [1]. The anomaly of virgin curve outside the envelope loop in an isothermal measurement of magnetization (or resistivity) as magnetic field is cycled has been extensively reported. The two different metastabilities that have accordingly been extensively investigated over the last twelve years correspond to (i) the measurement temperature lying in the region between supercooling and superheating spinodals [2]; and (ii) the measurement temperature lying below the kinetic arrest temperature of a disorder-broadened transition [3]. The anomaly corresponding to the first case is seen in a temperature window that is bounded from both below and above; the anomaly is non monotonic as a function of temperature; and the anomaly is seen in an isothermal measurement only if the measurement temperature is reached by warming (cooling) when the low-temperature phase has lower (higher) magnetization (see figure 3 of reference [2] ). Anomaly (ii) is a monotonic function of temperature, becoming more pronounced as temperature is lowered, and is attributed to glass-like arrested states formed by arresting a first order magnetic transition. Anomaly (i) has been reported in the temperature window between 120K and 140K in figure 1 of reference [2] for an intermetallic; while anomaly (ii) has been reported below 15K in the same material in figure 2 of reference [2]. Similarly, anomaly (i) has been reported in the temperature window between 125K and 215K in figure 2 of reference [4] for a shape memory alloy; while anomaly (ii) has been reported below 100K in the same material in figure 2 of reference [1]. The anomaly corresponding to the second case was first reported by Manekar et al [3], who invoked the Imry-Wortis work [5] to replace the supercooling, superheating, and the phase transition lines in (H,T) by bands, with lines within bands corresponding to regions with spatial extent corresponding to the correlation length.

Many more experimental manifestations of this incomplete transition have been reported, and the analogy with glass-like arrest was established by Chattopadhyay et al [6]. The Letter [7] being commented on, discusses in some detail "The presence of the virgin curve outside the envelope loop" seen in the isothermal M-H data in its figures 1c,1d, 2a, 3a to 3f. This anomaly is observed at 30K and below, and is an observation of the anomaly labeled (ii). This could easily be confirmed by measurements following the 'cooling and heating in unequal field (CHUF)' protocol (see [1] and references therein). The discussion starting in the right column of the second page and continuing into the right column of the third page is oblivious to this well understood distinction between the two distinct origins [2] of 'virgin curve outside the envelope loop'.

The authors of ref [7] have studied the same material in a subsequent paper[8], and recognized that this material exhibits kinetic arrest of the first-order ferrimagnetic to antiferromagnetic transition, resulting in the existence of a magnetic-glass state [8]. They have also reported measurements following the CHUF protocol as suggested above, without naming the protocol followed. They have concluded that "The observation of unequal magnetic behaviors when the sample is cooled with a field smaller/larger than the measuring field clearly indicates the kinetic arrest of the FM (FI) phase." This conclusion confirms that the paper [7] commented on was not justified in the open-ended conclusion stated in the last sentence of its abstract viz. "The field-induced irreversibility originates from an unusual first-order ferrimagnetic to antiferromagnetic transition".